# How We Ruined the Internet

## The information superhighway to hell was paved with good intentions

Many of the problems currently experienced by developers and users of the current ICT environment are attributable to the definition of the Internet as a datagram forwarding stovepipe.

Micah D. Beck

Dept. of Electrical Engineering and Computer Science, University of Tennessee, Knoxville, mbeck@utk.edu

Terry R. Moore

Innovative Computing Laboratory, University of Tennessee, Knoxville, tmoore@icl.utk.edu

At the end of the 19th century the logician C.S. Peirce coined the term "fallibilism" for the "… the doctrine that our knowledge is never absolute but always swims, as it were, in a continuum of uncertainty and of indeterminacy" [1]. In terms of scientific practice, this means we are obliged to reexamine the assumptions, the evidence, and the arguments for conclusions that subsequent experience has cast into doubt. In this paper we examine an assumption that underpinned the development of the Internet architecture, namely that a *loosely synchronous point-to-point* datagram delivery service could adequately meet the needs of all network applications, including those which deliver content and services to a mass audience at global scale. We examine how the inability of the Networking community to provide a public and affordable mechanism to support such *asynchronous point-to-multipoint* applications led to the development of private overlay infrastructure, namely CDNs and Cloud networks, whose architecture stands at odds with the Open Data Networking goals of the early Internet advocates. We argue that the contradiction between those initial goals and the monopolistic commercial imperatives of hypergiant overlay infrastructure operators is an important reason for the apparent contradiction posed by the negative impact of their most profitable applications (e.g., social media) and strategies (e.g., targeted advertisement). We propose that, following the prescription of Peirce, we can only resolve this contradiction by reconsidering some of our deeply held assumptions.

## 1 INTRODUCTION

The state of the Information and Communication Technology (ICT) environment— "the Internet", "the Web", and "Cloud Services"—has been the subject of growing waves of complaint, outrage, and distress for more than a decade. The problems that typically provoke these sentiments are not without precedent. For example, businesses in the last century routinely used mass media as a conduit for marketing propaganda and tried to create monolithic retail mechanisms to capture and manipulate customers. But the spread of Internet-powered social media and of targeted advertising enhanced by consumer surveillance has amplified and weaponized these strategies. This has made them much more worrisome and infuriating. Since the ICT hyper-giants are leading players in the AI revolution, it seems unlikely that these trends will abate anytime soon.

Some commentators who lived through the development and growth of Internet-connected distributed systems have found this state of affairs especially perplexing. For them, there is a stark disconnect between the intentions and goals of the early champions of the Internet as the foundation for an "Open Data Network (ODN)" [2] and the disturbing aspects of the ICT landscape we live in today. There is a striking and unexplained contrast between our early expectations and aspirations and the destructive social and political results we now see. This broken vision of an ICT-driven renaissance is puzzling enough to be characterized as a “paradox” [3].

Explanations of this seeming paradox vary. The most common approach holds that the vision has been abandoned in favor of the business models which support a global system of infrastructure and services. Such accounts tend to focus on how corporate greed and unscrupulous behavior fueled this transformation [4]. But in this paper we offer a more technical explanation for many of the problems of the current ICT environment. We point to the attempt to realize the global ODN vision of the 1990’s by building on the Internet’s end-to-end paradigm of communication infrastructure as the key element. As we describe below, the result is the creation of a *stovepiped* communication infrastructure.

*That strategy was not able to address one of that paradigm’s fundamental limitations, namely the lack of in-network support for asynchronous point-to-multipoint services.* The need for auxiliary mechanisms to overcome this deficit created a ripe opportunity for the twenty-first century's hegemonic business models. These have had lamentable social, political, and economic consequences. We do not rule out avarice or unbridled ambition as important factors, nor do we downplay the importance of addressing them. However as members of the Computer Science and Engineering community we have a particular ethical responsibility to face the underlying technical factors that made such strategies necessary.

## 2 A TALE OF TWO “INTERNETS”

We begin by addressing a terminological difficulty in discussing the history and current state of the ICT environment. What the term “The Internet” designates has changed over time, so that what it names in the technical networking community is often different from what it refers to in society at large. This ambiguity has in led to a misunderstandinng about the nature of the infrastructure that supports modern Web applications at scale. In order to avoid such confusion, we have assigned different names to these distinct referents: the Internet* and Internet++:

- ***Internet**** — We call the communication network that embodies most closely the original Internet architecture *the Internet**. Also referred to as the Internet Protocol Suite, its most well-known elements are IP (the Internet Protocol) and TCP (the Transmission Control Protocol). Components of the Internet* that are not visible to end users include ICMP (the Internet Control Management Protocol) and internal and external (global) routing protocols such as OSPF (Open Shortest Path First), RIP (Routing Information Protocol), and BGP (Border Gateway Protocol). Internet* services comprise Layers 3 (Network) and 4 (Transport) of the Internet Protocol Stack [5]. The asterisk is a reminder to the reader that the Internet* is not Internet++.
- **Internet++** — We refer to the environment of applications and other services that are used today by hundreds of millions of users and businesses worldwide as Internet++. This includes foundational services such as remote login and email, as well as ubiquitous facilities such as the World Wide Web, Web search, and social media. It also includes the additional infrastructures (e.g., Content Delivery Networks and Cloud data centers) that have grown up to support today's more general application requirements.

When the developers and early advocates of the Internet* argued for its adoption as the basis for America's National Information Infrastructure (NII) (and implicitly for the whole world's information infrastructure), their vision was something quite different from Internet++. They envisioned an *Open Data Network* (ODN) that achieved four different goals (from the 1994 National Research Council report "Realizing the Information Future" [2], p. 44):

- *Open to users: It does not force users into closed groups or deny access to any sectors of society, but permits universal connectivity, as does the telephone system.*
- *Open to service providers: It provides an open and accessible environment for competing commercial or intellectual interests. For example, it does not preclude competitive access for information providers.*
- *Open to network providers: It makes it possible for any network provider to meet the necessary requirements to attach and become a part of the aggregate of interconnected networks.*
- *Open to change: It permits the introduction of new applications and services over time.… It also permits the introduction of new transmission, switching, and control technologies as these become available in the future.*

But adopting the Internet* as the foundation for an ODN has produced instead Internet++, a global information infrastructure that has properties destructive to some of the core values informing this ODN vision, which many still hold dear. Our hypothesis is that some technical features of the Internet* helped to cause this paradoxical outcome.

Admittedly the considerations underlying this hypothesis are not purely technical. They also draw on historical records of the intentions and expectations of early Internet* advocates and formal principles of system design. We suggest that some of the early assumptions of the Internet* architecture may need to be reconsidered if we are to regain momentum toward those early goals. It is noteworthy that "thought leaders" in the fields of Networking and Distributed Services assert that no new infrastructure can trade off strong properties such as low latency bounds and high availability. They rule out such developments even as a means reach the desirable goal of affordable universal service for the world's entire population. Instead, the community endorses policies such as subsidized broadband which are understood to be insufficient to meet the goals of universal service. However, such policies *do* conveniently direct large sums of money to the employers and benefactors of those who promote the status quo.

### 2.1 The Internet Stovepipe

Stovepipes are "*... systems procured and developed to solve a specific problem, characterized by a limited focus and functionality, and containing data that cannot be easily shared with other systems.*" ([6], p. 133) In system architecture the term "stovepipe" refers to a collection of services implemented within each level of a layered system architecture. The services which comprise the stovepipe communicate only with other elements of that collection to export a service interface that is restricted to a specific scope. A stovepipe thus defines a vertical information conduit or "slice" of the layered system (Figure 1).

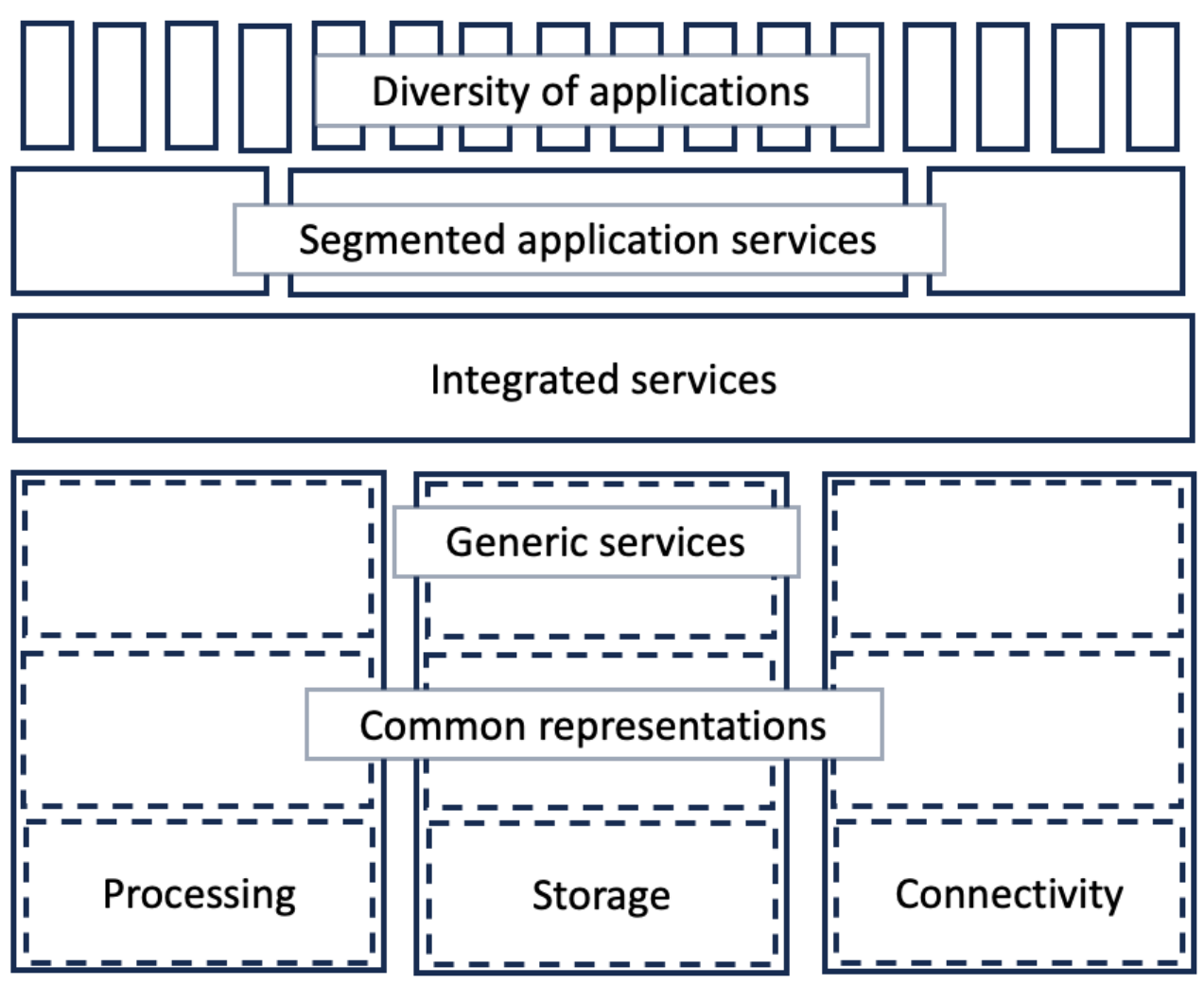

Figure 1: The Processing, Storage and Connectivity stovepipes at the base of the distributed application services stack. Adapated from Messerschmidt and Szyperski. [7]

Accounts of ICT infrastructure architecture identify three fundamental stovepipes, namely Storage, Networking and Computation ([7], p. 215). Data that resides in secondary storage, for example, typically occupies different physical resources from data that moves through wide area networks. These separate facilities are managed in vertically isolated subsystems known as "stacks". The separate Storage, Networking, and Computation stacks generally exchange data or interact with process management components of the operating system only at the highest level, through application interfaces. Since the applications these systems support need to combine the use of all three of these physical resources, the fact that they have to repeatedly access three separate interfaces atop three separate stovepipes can be a source of inflexibility and inefficiency [8; 9].

### 2.2 The Internet* Stovepiped Spanning Layer

In a layered system, a *spanning layer* is a distinguished set of system elements that partitions the system's software stack vertically, with applications that are created using this interface above, and the *lower level services* required to support it below (Figure 2) [10]. The purpose of the spanning layer is to enable *interoperability* among different application implementations. This that they can use different supporting services without change to application code. Such interoperability is best served if the layer separating applications from the supporting services is *strict*, meaning that there is no direct access by applications to those services. A spanning layer is *stovepiped* if it offers to applications only a service that addresses a narrow/restricted type of functionality, which is insufficient to meet all application requirements. For example, like every digital application or service, an Internet* router (or intermediate node) makes coordinated use of data transfer, persistence, and processing resources; but the latter two primitive services are encapsulated within the implementation of the Internet Protocol Suite, which defines and exports a simple datagram forwarding service.

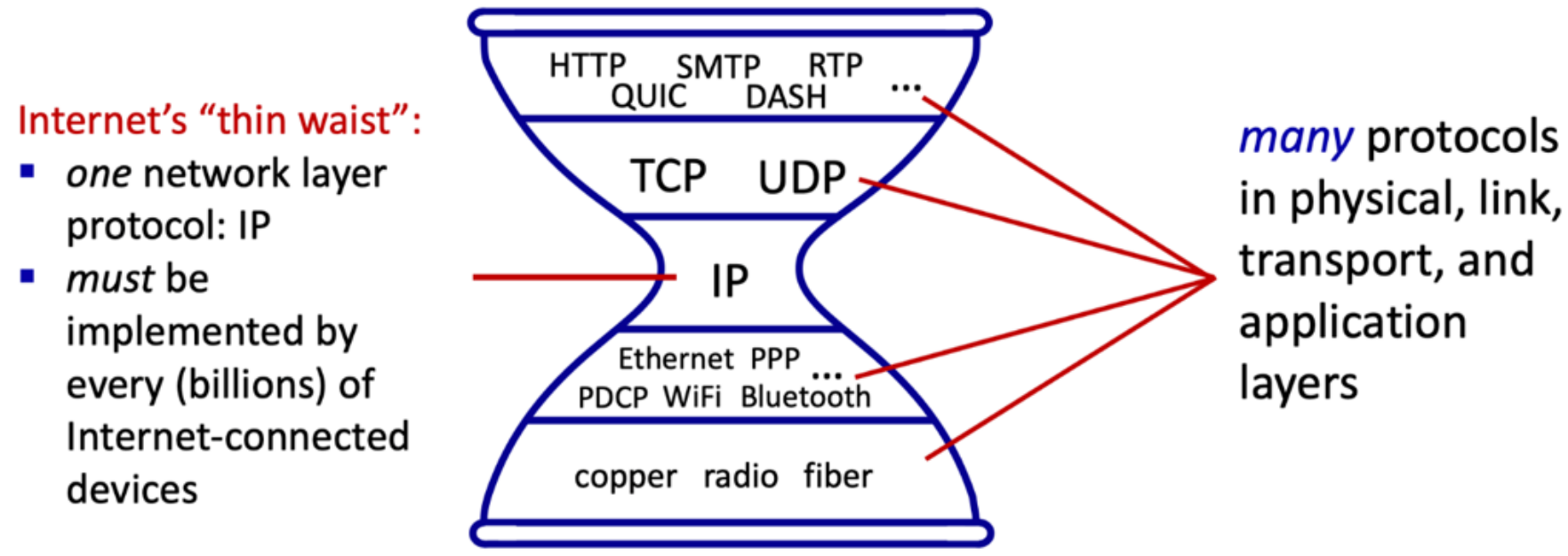

Figure 2: The Internet Hourglass stack with the Internet Protocol spanning layer at its waist [11].

A spanning layer is often used to define a standard that members of some community adopt to enable interoperability. What happens when a stovepiped spanning layer is defined as a community standard? One result, already noted, is that users are likely to confront problems of inflexibility and inefficiency in combining this standard with other resources that are required to address their varied application requirements. But there is potentially a more serious issue.

This issue can arise when a spanning layer is adopted as the standard interface to the ICT infrastructure that distributed systems require, namely the Internet*. Application requirements ultimately determine the resources that must be used. And how application implementers get access to those resources can in turn determine the structure of the infrastructure that supports them. Attempting to set standards that constrain these forces can only succeed as long as the economic or policy power of the standard setters is greater than that of the application communities. If the application communities gain the upper hand, as in the case of Internet++, standards that constrain application communities will be circumvented or ignored.

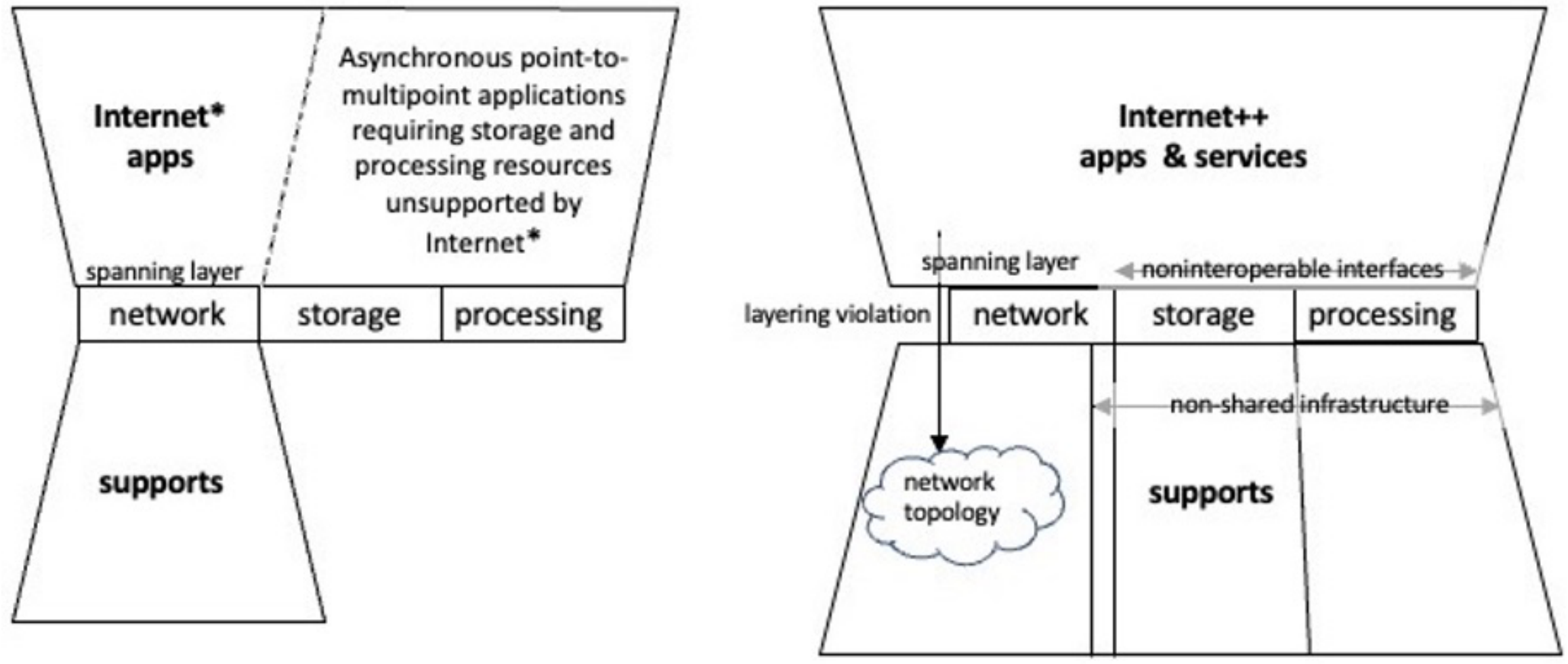

Figure 3: The Internet* is used as a loosely synchronous point-to-point communication service that does not itself provide the storage and processing services that many applications (e.g. mass media) require. Internet++ has two basic ways of working around this problem: layering violations and private infrastructure.

As shown in Figure 3 a stovepiped communication spanning layers that does not to support application requirements for distributed storage and processing will tend to provoke two possible responses:

1. Some applications will violate the strictness of the spanning layer, reaching beneath it to directly access underlying services that are supposed be encapsulated behind its specialized interface; and/or
2. Non-shared infrastructure, not constrained by the spanning layer, will be constructed to support the needs of particular classes of applications.

Option 1 can result in a system that does not have full interoperability. This tendency can be inconvenient but, if limited, can still be managed. In fact, the ability to violate the strictness of a spanning layer is sometimes even held up as a virtue [12], because it enables applications to innovate faster than the definition of the spanning layer can evolve. Option 2 is also sometimes held up as a virtue because it allows the limited functionality of the stovepipe to be augmented by building up auxiliary infrastructure, enabling both innovation and additional investment by users. However the level of investment in such auxiliary infrastructure, and the economic power of those who invest in it, can grow to overshadow the stovepiped infrastructure. In such cases the results may be, and to some degree have been, completely at odds with the intentions and expectations of the original advocates of the Internet*.

### 2.3 Asynchronous Point-to-Multipoint Applications At Scale

To understand the forces that drove the Internet* to evolve into Internet++, the key thing to notice is that there is a fundamental disconnect between the requirements of many of the most popular and profitable Internet++ applications (mass media and public utilities) and the nature of the network service provided by the Internet* (Figure 1). Specifically, many if not most of the former require *asynchronous* *point-to-multipoint data and service delivery*, while the **only** ubiquitously deployed communication mechanism of Internet* is *a loosely synchronous one based on point-to-point datagram delivery*.

The structure of simple file and object distribution applications is naturally asynchronous and point-to-multipoint: a file is published and advertised to users, who then request delivery of identical content at times and locations of their own choosing. To address this requirement using the point-to-point synchronous functionality of Internet*, it has to be decomposed into two phases, in the form of "iterated unicast":

1. A file is published and stored within the file system of a server; this server is not a component of Internet*, but a network endpoint that uses Internet* as a medium. The server then listens for asynchronous TCP connection requests.
2. To retrieve a given file, each client initiates an independent point-to-point TCP session, resulting in synchronous communication during which identical file contents are delivered via an application protocol such as FTP (the File Transfer Protocol) or HTTP (Hypertext Transport Protocol).

By distinguishing between the structure of file delivery *as an application* and the unicast nature of its Internet* *implementation*, one can see that the auxiliary server infrastructure which asynchronous point-to-multipoint applications require falls outside the network itself, and thus lies beyond the control of the Internet*'s implementers and advocates. As the system scales up, this divide inevitably widens because implementing point-to-multipoint applications using iterated unicast introduces inefficiency in the use of both network and server resources. The pressure on these resources increases linearly with the number of simultaneous client requests, and also with the size of the files to be distributed. At scale, the resources of the server and/or the network to which it connects tend toward exhaustion of local resources, resulting in inadequate responsiveness, inability to connect, or server instability. This "hotspot" problem (the so-called "Slashdot effect") was well known to early users of FTP. It exposes the inadequacy of purely point-to-point communication in supporting many critical applications at scale.

The principles of the Internet* lead to a very different response to the hotspot problem than that which has been adopted in Internet++, and which we will discuss in Section 3. The canonical Internet* response is to advocate for the use of an efficient in-network solution, such as IP multicast, which delivers datagrams from a sender to multiple receivers using tree-structured forwarding. However IP multicast fails to meet the needs of asynchronous point-to-multipoint applications in a number of important respects: it is synchronous, it does not support error correction, and it is more complex to implement and use [13]. These issues with IP multicast are not incidental to a particular protocol design; they are inherent to working within a successful point-to-point communication network which is growing explosively at a global scale. All of these issues could be addressed through the management of more persistent data state within the Internet* network. But as we discuss in Section 3.2, such strategies come into direct conflict with the "fast path" optimizations required for routers implementing unicast data forwarding at scale.

**Distributed unicast drives CDN infrastructure growth**

Consider a computer network consisting of m servers and n clients connected through a graph of intermediate nodes connected in some topology. The content distribution problem is to store a set of files on server nodes and to deliver a copy to any client when they request it.

If the data is stored on a single server, the network traffic generated by n file requests is proportional to n*p where p is the average length of a path from source to client, with the load at the source also proportional to n. When directed at a single server this load can cause a "hotspot," i.e., overloading of local network and/or processing resources (see Figure 4a). If the network's intermediate nodes are allowed to store file contents and to implement appropriate policies, the content distribution problem has a number of efficient solutions which involve passing files over a tree (Figure 4b). While the number of tree edges is dependent on the network topology, total traffic is typically proportional to n, and the load at any server is bounded by a constant.

The problem of server and network overload can be mitigated without constructing a distribution tree through distributed unicast: replicating the input file on m servers and distributing load evenly, thereby diminishing the network and processor load local to each server by a factor of 1/m. Content Delivery Networks (CDNs) distribute files efficiently across their own servers and then use distributed unicast to reach clients (Figure 4c). This results in the average load on any server being proportional to the number of servers, but still generates total unicast traffic proportional to n*p', where p' is the average (topologically optimized) path length from a CDN node to a client (Figure 5).

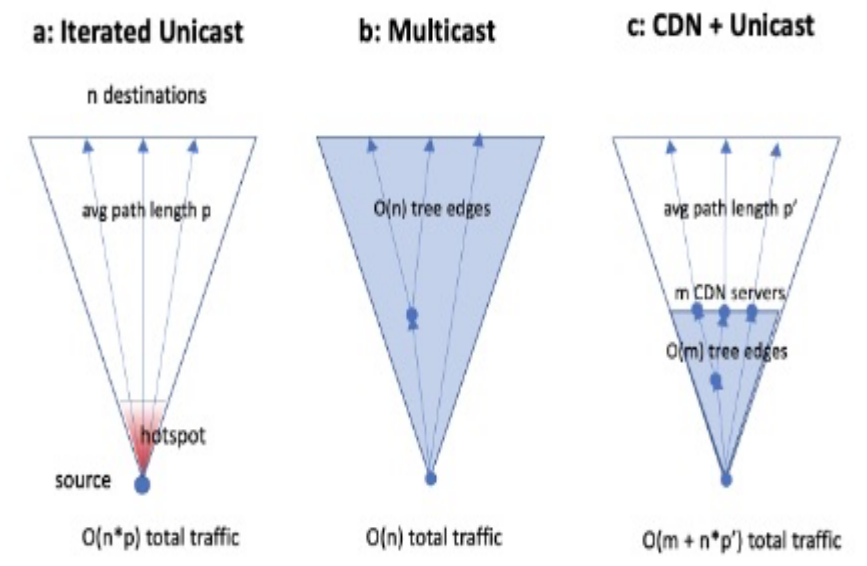

Figure 4: Unicast, Multicsast and CDN

Figure 5: CDN servers per PB of Global traffic

Keeping up with the explosive growth in total Internet traffic (from 122 EB/mo in 2017 to 293 EB/mo in 2022) thus requires an increase in the number of servers deployed by one of the largest global CDNs (from 200K in 2017 to 350K in 2022[1]). This keeps the number of servers roughly proportional to the total volume of data delivered to clients, limiting the load per server. In fact the number of ration of server nodes deployed to support global total traffic has stayed roughly constant in recent years (see Figure 5).

[1] These statistics, drawn from public industry estimates, are not authoritative.

## 3 THE PARADOXICAL PATH FROM THE INTERNET* TO INTERNET++

When we look back to the early 1990's and the idea of creating an Open Data Network as a global information infrastructure, perhaps the strongest argument for using the Internet* as its foundation relied on the fact that the stateless model of point-to-point datagram delivery that the Internet* offered had already demonstrated its global scalability. In 1994, the Internet* was a network of networks consisting "... of approximately 20,000 registered networks, some 2 million host computers, and 15 million users," and it was offering people around the world unprecedented international communication and publication capabilities [2]. Moreover, the experience of building the Internet* had provided convincing evidence that its defining service—unicast datagram delivery—could be deployed at scale with the kind of universality necessary to reach everyone everywhere. The unanswered question was whether that service could continue to meet the application requirements and other objectives of the ODN vision as the system scaled up. These included decentralized management and the ability of the Internet* to evolve. Contrary to the next three decades of networking community orthodoxy, the answer to that question turned out to be "No".

As the global network of networks grew, the limitations of the Internet*'s core service had consequences both outside and inside the system it created. Externally, the need to provide asynchronous point-to-multipoint service led to the emergence of CDNs and Cloud services, which included proprietary overlay routing mechanisms that repurposed Internet*'s support infrastructure (e.g., DNS and BGP). Internally, the effort to scale up distributed unicast to keep up with the growth of the network required router optimizations—the data "fastpath"—that ossified Internet* itself at the Internetworking Layer [14], which is the IP spanning layer (Layer 3 of the Internet* stack). We treat each of these issues in turn below.

### 3.1 Why applications required another network (Internet++)

The ability to provide asynchronous point-to-multipoint services is a fundamental requirement of a broad and important class of network applications. Given that the only ubiquitous service offered by the Internet* is unicast datagram delivery, those aiming to provide asynchronous point-to-multipoint services, absent any mechanism to constrain them, developed and relied on a sequence of work-arounds. Since these alternative measures were initially implemented in overlay using only application layer mechanisms, such work-arounds were once (and sometimes still are) viewed as applications of networking, rather than as alternative networks with their own architectures. These mechanisms use the components (subnetworks) which comprise Internet* as building blocks and public facing on-ramps. But a little reflection makes it clear that they have evolved to implement their own global topologies and policy mechanisms.

The point-to-multipoint nature of file distribution was expressed from an early date through the use of "FTP mirroring". This strategy involves a single file being copied to multiple servers, which then become alternative sources from which clients can request distribution using iterated unicast. In the early days of the Internet*, FTP servers were not profit centers. Consequently, the purpose of investing in them, and in organizing them into collections of FTP "mirrors", was viewed primarily as a means of improving and scaling file distribution. It was not foreseen that such servers would be constructed and operated specifically to implement service replication, or that they would become costly utilities that had to be supported by ample streams of income. Yet clearly, Content Delivery Networks (CDNs) and then Cloud networks have evolved to serve that purpose and, in so doing, have become new centers of power over the distributed environment.

Given that context, one way to see that CDNs are genuine networks is by looking at the way they route the data that flows through them. In the Internet* architecture, the choice of a path for a datagram through the network is governed by routing protocols. But antecedent to this routing of datagrams, a CDN that builds on top of Internet* has to make a higher level choice of which replica server a client request should be directed to. Making such choices to optimize network traffic and server load requires knowledge and management of the underlying resource topology, which makes it a kind of coarse grained routing. Such routing requires that CDNs be implemented so as to reach below the Internet Protocol in order to access and even control the underlying topology of the network.

Leveraging the Internet* Domain Name Service (DNS) plays a key role in this strategy. The DNS implements a mapping of domain names to sets of IP addresses. The set of IP addresses bound to a domain name is returned in an arbitrary order which carries no information or preference between members of the set [15]. Early CDNs tried to implement policy-based server selection without usurping the intended function of the DNS, but such application-level strategies were ultimately inadequate.

The solution that has been widely adopted is a nonstandard modification of the behavior of the DNS. A DNS server can get access to Link Layer (Layer 2) topology by communicating directly with Border Gateway Protocol (BGP) routers. They can then use this information to choose with some topological accuracy the IP address to which each client request should best be directed. By returning only one IP address, a CDN's private DNS server can impose its choice on each client [16].

This strategy makes the DNS the central mechanism for the direction of client requests to alternative servers, which is a form of routing used by CDNs. Modern CDNs not only access internal BGP topology information, they sometimes take a hand in determining the placement of servers within the Layer 2 network, even shaping the topology through contractual arrangements with Internet Service Providers. Thus CDNs create their own networks by circumventing and/or repurposing the components of the Internet*.

### 3.2 How distributed unicast led to network ossification

As discussed above (see inset text), keeping up with the explosive growth in the user community and the demand for increasingly large files requires a continuous investment in CDN server and network capacity to implement *distributed unicast*. While distributed unicast cannot match the near-optimal efficiency of fully tree-based point-to-multipoint communication mechanisms, it has other strengths. These include leveraging the simplicity and ubiquitous deployment of unicast datagram delivery service which supports important functionality such as congestion control and data integrity through retransmission (e.g., TCP and QUIC). Moreover, it makes it possible to dynamically generate responses to user requests, adapting them to the identity of the end user or the context of each request.

The failure to deploy a shared solution to the content delivery problem which is efficient in terms of total traffic has resulted in the growth of a private infrastructure proportional in size and greater in economic power to the Internet* communication infrastructure. In addition, the burden of supporting massively growing unicast traffic necessitates the deployment of Internet Protocol forwarding nodes which have the highest possible capacity and the lowest possible latency. These requirements make the extension of those intermediate nodes to also support the management of stored data much harder, if not impossible. Thus the adaptation of Internet* to the pressure created by iterated unicast has the effect of ruling out the adoption of more efficient solutions to the content distribution problem.

The forwarding of a datagram arriving on the input port of an intermediate node to a single outbound port can be implemented using hardware acceleration—the "fast path"—which boosts peak performance to otherwise unattainable levels. By implementing datagram forwarding in hardware to create an optimized fast path, the excess traffic that CDN servers generate through their use of distributed unicast can be handled by investment in server and router infrastructure. Moreover, the incredible performance level attained by modern backbone routers has enabled new applications, such as immersive telepresence, that demand both high bandwidth and extremely low latency. These applications are expected to serve a smaller audience but to be more profitable than current Internet++ applications. However, they place ever greater demands on backbone routers and have the impact of constraining innovation and generality in the services they provide [14].

## 4 LEVERAGING THE FAST PATH: ON-DEMAND STREAMING AND TELEPRESENCE

The Internet* architecture is nominally store-and-forward in design [17], and thus asynchronous. However in practice it has always been used as a loosely synchronous network. A fully asynchronous network (having no bound on latency or loss of individual packets) cannot necessarily support some demanding applications that were part of the original global ODN vision. These include real-time video streaming, telepresence (e.g., audiovisual conferencing) or remote control. The idea that the Internet* architecture would support such low latency applications implicitly assumes an implementation that is statistically synchronous, meaning that there are bounds on latency, jitter and loss over short periods of time, as well as continuous service availability over long periods. Presumably the assumption that low latency would be a characteristic of a well-designed, well implemented, and competently operated global network seemed reasonable during the period of early Internet* growth. At that time, the underlying networks were supported almost exclusively by wired infrastructure and technological advances in transmission were proceeding at a breakneck pace. But as noted above, "fast path" acceleration in Internet* backbone routers is now a necessary condition for making good on that assumption.

What no one at that time seemed to recognize, however, was that building in the assumption of low latency represents an implicit *strengthening* of the fundamental communication service; the RFCs defining IP connectivity specify no such characteristics (the IPv4 Time-To-Live header field is used to count forwarding hops). The Hourglass Theorem [18] predicts that such strengthening of the common communication service comes at a cost: the scope of applications that can be supported will increase—now we can have interactive telepresence—but at the cost of ruling out some possible implementation strategies. In particular, if the alternatives that are ruled out include some that are low in cost or high in ease of implementation, the digital divide in broadband is likely to remain wide or even grow, as has happened [19].

The reality of the global communication network is that it now targets a much broader public than was originally envisioned, with the largest and fastest growing end user population supported by less capable wireless connectivity to mobile phones [20; 21]. High performance services such as video streaming and telepresence not only place high requirements on the communication network, they are also in high demand by consumers in the parts of the world that have the best communication infrastructure and the highest levels of disposable income. This continues to draw the efforts of the builders of CDN and Cloud infrastructure away from less synchronous services to these highly profitable ones that require extremely low latency.

The commercial imperative to support real time streaming and telepresence is now implicitly included in the definition of acceptable broadband service. This fact drives up the cost of connectivity to end users, including those whose critical need is for services that *can* be delivered asynchronously. Deploying only infrastructure designed to = support low latency imposes high costs while increasing network instability, or unreliability in the quality of telepresence and lack of universal access. This choice of strategy is subsidized by large corporations and the governments of wealthy nations. However, it is noteworthy that, in the implementation of services such as streaming video, large application vendors take advantage of buffering outside the network to support their most profitable user communities. It is only the use of less synchronous mechanisms within the public network to support critical applications for those least able to pay that is forbidden.

For example, although real-time media streaming requires high bandwidth, low latency communication from source to receiver, services such as Netflix do not rely on highly demanding real time connectivity. Instead, they collaborate with edge resources (in the end user device or a streaming appliance) to buffer the media stream, allowing much looser bounds on delay and loss. In addition, by placing replicas in edge networks of the most valued customer base, the strategy of replicating source files to deal with the inefficiency of iterated unicast has been adapted to minimize the path that must be traversed to reach the end user. This means that content stored in CDNs need only traverse the "last mile" between a well-placed replica server and the most valuable end user using shared public infrastructure. The internal distribution of content to CDN servers is under the control of content vendors or their business partners and so must be supported by their business model.

## 5 THE TAIL WAGS THE DOG: PAYING THE BILL FOR ICT INFRASTRUCTURE

So far we have argued that Internet++ evolved as a private response to the inability of the publicly-minded network research community to evolve the Internet* to meet the need for asynchronous point-to-multipoint services and content distribution. This outcome should not be surprising. If the public infrastructure is unable to meet the evolving needs and aspirations of the application community, we should expect that profit-seeking interests will be incentivized to construct additional infrastructure to satisfy these demands. Indeed, the buildout of CDN infrastructure by companies such as Akamai provided a model for profitable investment in private distributed storage and processing resources to support, at scale, asynchronous point-to-multipoint applications and services. Following a similar path, several major cloud data centers also emerged from specific application domains: Web search (Google), online shopping (Amazon), social media (Facebook) and streaming media (Apple). Meanwhile, Microsoft went into the business of serving generic applications. Cloud services defined a new paradigm for the creation of applications with huge user communities. The private nature of such CDN and Cloud infrastructure obviated the need for infrastructure to be usable by all applications, in particular, by those which are not supported by large income streams generated by for-profit business models.

Private ownership and centralized control of such essential infrastructure in turn presents an opportunity to maximize the profitability of online services by taking advantage of features that are not supported by public networks, even when such strategies are viewed as harmful by the user community. In the case of Internet++, such features include the following: the ability to surveil the activity of users and to collect and resell their personal data; the ability to utilize that data to create a business model that promises much more accurately targeted and effective advertising; and the ability to maximize the profit such models generate by creating engagement-maximizing algorithms that increase the amount of time and attention users devote to online applications, such as social media and gaming. The lack of any notion of utility other than profit clearly encourages applications that are increasingly addictive, often with apparent indifference to the well-being of users or the health of public discourse.

An alternative vision of a more benign Open Data Network, built on a stateless model of unicast datagram delivery, clearly seemed plausible in 1994. Abandoning this vision, Internet++ has appropriated Internet* mechanisms for its own ends, repurposing its components to support an altogether different kind of network. The resulting contradictions between the values that inspired the "Internet revolution" and the reality that has evolved from them are bound to seem paradoxical to many now:

- ODN was supposed to be open to users, but in the context of Internet++ the oft-stated commitment to universal network connectivity belies reality. On the contrary, the broadband requirement of low latency to support increasingly demanding real time interactivity, makes universal connectivity increasingly difficult, expensive, and unlikely [19].
- ODN was supposed to be open to service providers, enabling competitive access to information services, and open to network providers. It was supposed to enable any provider to attach and become part of the aggregate system. But supplying an application or a service with access to Internet++ is not a simple matter of attaching a single server (or even a cluster) to the IP backbone. It requires the deployment and connection of large numbers of CDN replica servers, or a smaller number of Cloud data centers, throughout the network. This path is not open and accessible for competition, but has instead led to the emergence of a cartel of Internet++ “hypergiants” that use engineering dominance and monopolistic market power to concentrate control over application provisioning. Access is restricted to information and service providers who can pay their fees and are willing to obey their rules. (See for example Apple’s control over its IOS App Store.)
- ODN was supposed to be open to change, including the introduction of new transmission, switching, and control technologies. Yet as a recent article points out, "The functions performed at the internet layer of the protocol stack are no different from those of 25 years ago," as shown by the fact that "... IP Mobility, Multicast and IP Security (IPSec) are largely Internet layer technologies that have failed to gain significant levels of traction in the marketplace of the public Internet" [22]. Thus, the IP network that has been assimilated by Internet++ and ossified by the need to support iterated unicast. The only network innovations that are now considered feasible are those that are compatible with routers supporting low latency, high bandwidth point-to-point datagram delivery at massive scale. This has precluded the introduction of new approaches to network connectivity such as asynchronous point-to-multipoint data delivery, much less in-network storage and processing.

These outcomes contradict networking community orthodoxy, which for the last thirty years has held that a point-to-point, quasi-synchronous communication service can provide an adequate foundation, at scale, for an ODN that supports most if not all kinds of distributed applications. This was the assumption in 1994, when a National Information Infrastructure based on the Internet* was envisioned for the United States; a review of its anticipated applications—email, education, text archives, entertainment, financial services, data storage, and news ([2], p. 34)—clearly shows as much. Since these applications rely on asynchronous point-to-multipoint content and service distribution, in order to deliver them at scale, the next generation pioneers of Internet++ commandeered and repurposed the mechanisms and infrastructure of the Internet*, augmenting and optimizing as necessary with private extensions and alternatives. As we have argued, however, the networking community's unicast datagram dogma has turned out to be false, not because Internet++ hasn't actually delivered all these applications and more, but because overcoming the inadequacies of the point-to-point datagram forwarding has, in the process, forced us to sacrifice all of Open Data Networking's defining goals.

## 6 CONCLUSION: GETTING BEYOND THE INTERNET STOVEPIPE

We began our discussion of the Internet stovepipe hypothesis by differentiating the Internet* from the Internet++; and we proceeded to explain how the technical limitations of the former led, absent any successful intervention by the Network Research community, to the latter. The result has been an application environment that continues to be called “the Internet”, but which has largely abandoned the goals of Open Data Networking in favor of expensive strategies that primarily serve well financed commercial interests or highly subsidized governmental ones. Applying this legacy name to today’s global information infrastructure obfuscates this transformation and its exclusionary and hegemonic nature; and it leaves us wandering among seeming paradoxes, unable to find a path forward to the recovery of the fundamental values to which the networking community once aspired.

Where can we look for such a path forward? Obviously any thorough discussion of how we might escape today’s frustrating socio-technological cul-de-sac must go well beyond the confines of this paper. But while it is certainly more comfortable to blame corporate greed or governmental failures for the darker consequences of the Internet revolution, it is hard to see how the situation can be substantially improved until the Network Research community takes responsibility for its acquiescence in a series of work-arounds to the inadequacies of the Internet* (so-called “barnacles” [23]) that have turned out to be fatal to the goals of Open Data Networking.

At the end of the 19$^{th}$ century the logician C.S. Peirce coined the term “fallibilism” for the “… the doctrine that our knowledge is never absolute but always swims, as it were, in a continuum of uncertainty and of indeterminacy” [1]. In terms of scientific practice, this means we are obliged to reexamine the assumptions, the evidence, and the arguments for conclusions that subsequent experience has cast into doubt. However, path dependence due to material incentives, psychological ossification, or a sense of our own potential culpability can make reconsideration of this kind seem impossible. And this reluctance is made even stronger when the theories that need to be questioned lie at the foundation of technologies and infrastructures that billions of people depend on daily, even though we see that these creations now threaten our physical safety or undermine our most cherished values. Such aversion to the admission of error is said to have led Max Planck to observe that science only advances one funeral at a time. But just as Presidents Jefferson and Lincoln felt a need to point out that the U.S. constitution is not a suicide pact [24], we should likewise acknowledge that pursuit of friction-free wide area ICT (or human-seeming AI) is not a suicide pact either. If we submit to entrenched orthodoxy rather than fulfill to our duties as scientists and engineers, then no one else will save us from the consequences of our own creations.